LAUR - 06- 0563

# Numerical simulations on cleaning the neutron trap for measuring the neutron lifetime

**Vyacheslav N. Gorshkov and Gennady P. Berman**

Complex Systems Group, T-13, Los Alamos National Laboratory, Los Alamos, NM 87545

## Abstract

We present the results of numerical simulations of the dynamical behavior of trajectories of ultra cold neutrons (UCN) in a magnetic trap. The main goal of our simulations was to optimize the trap parameters in order to minimize the characteristic times for removing from the trap those untrapped neutrons with relatively long escape times (cleaning the trap). Our results demonstrate that, by a proper choice of the trap parameters cleaning times can be reduced to 15 sec, or even less. Many other dynamical characteristics of the neutrons in the trap, including the conservation of the adiabatic invariant which characterizes the orientation of the magnetic moment along the local magnetic field, are also discussed.

## 1. The geometry of the trap, the main parameters and relations

The geometry of the trap (suggested by David Bowman, *et al.*, LANL) is presented in two figures below. (See also [1] for discussions of a chaotic cleaning in a vacuum quadrupole trap.)

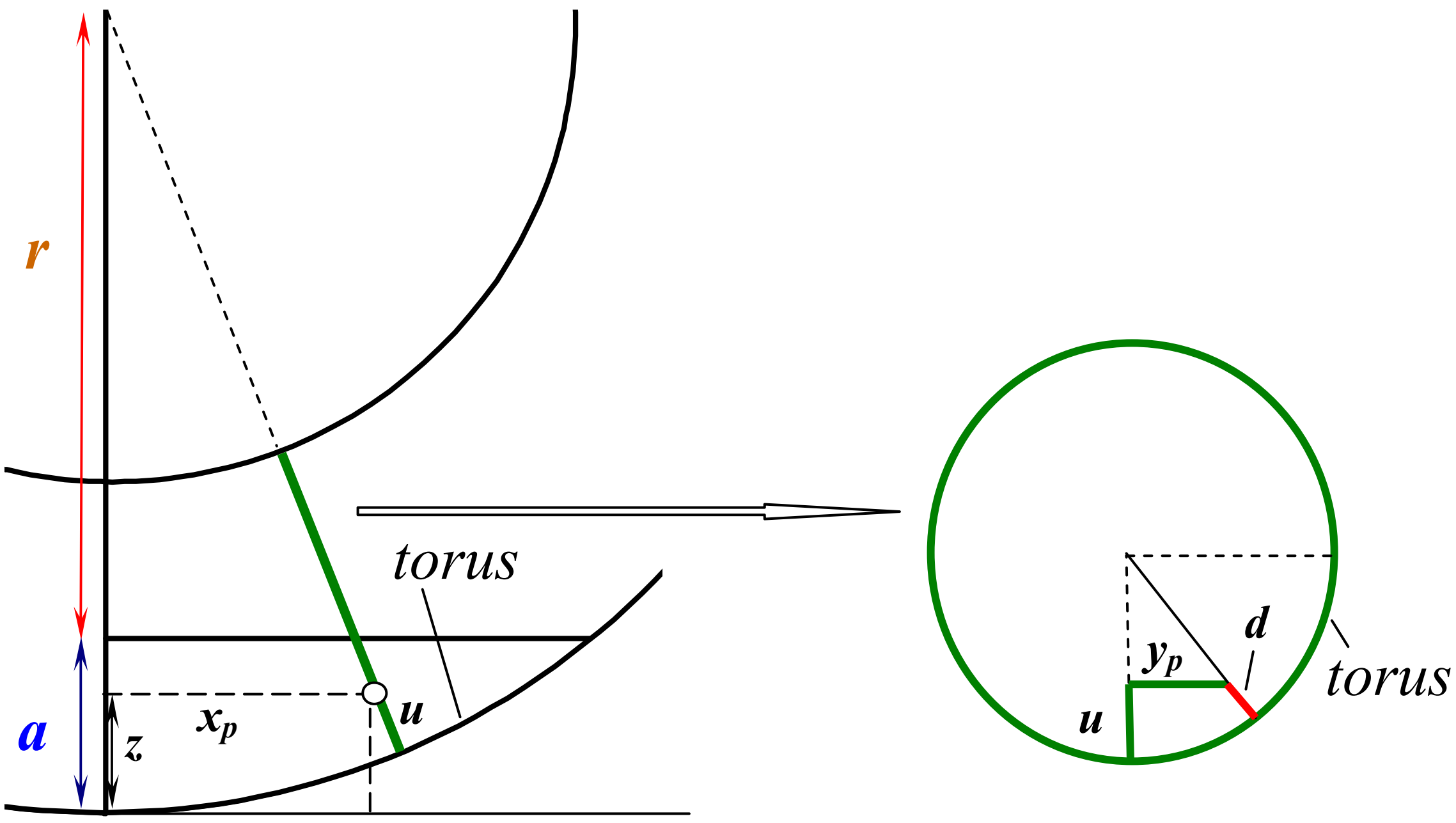

**Building blocks of the magnetic trap.**

The relations between the parameters of the trap are:

$$x_p = \begin{cases} 0, & |x| < L/2 \\ |x| - L/2, & |x| \ge L/2 \end{cases}, \quad y_p = \begin{cases} 0, & |y| < w/2 \\ |y| - w/2, & |y| \ge w/2 \end{cases}$$

$$ra \equiv r + a$$

$$u = ra - \sqrt{(ra-z)^2 + x_p{}^2}, \quad \sqrt{(ra-z)^2 + x_p{}^2} = ra - u \equiv raMu - \text{name of the variable}$$

$$d = a - \sqrt{(a-u)^2 + y_p{}^2}, \quad \sqrt{(a-u)^2 + y_p{}^2} = a - d \equiv aMd - \text{name of the variable}$$

$$\frac{\partial d}{\partial x} = -\frac{(a-u)}{\sqrt{(a-u)^2 + y_p{}^2}} \frac{x_p sign(x)}{\sqrt{(ra-z)^2 + x_p{}^2}} = -\frac{(a-u)}{aMd} \frac{x_p sign(x)}{raMu}, \quad sign(x) = \begin{cases} 1 \; if \; x \ge 0 \\ -1 \; if \; x < 0 \end{cases}$$

$$\frac{\partial d}{\partial z} = \frac{(a-u)}{\sqrt{(a-u)^2 + y_p{}^2}} \frac{(ra-z)}{\sqrt{(ra-z)^2 + x_p{}^2}} = \frac{(a-u)}{aMd} \frac{(ra-z)}{raMu}$$

$$\frac{\partial d}{\partial y} = -\frac{y_p sign(y)}{\sqrt{(a-u)^2 + y_p{}^2}} = -\frac{y_p}{aMd} sign(y)$$

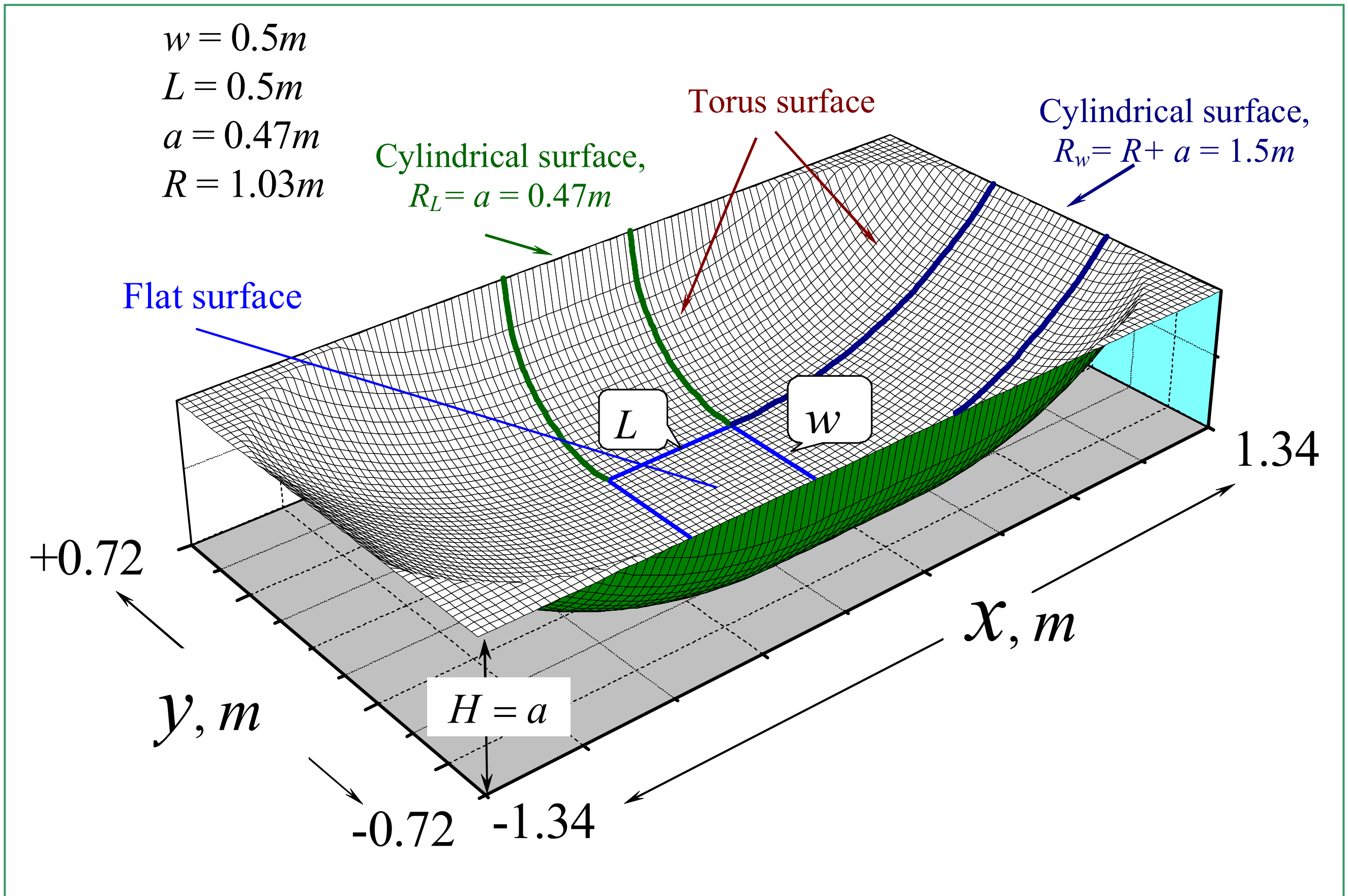

Geometry of the magnetic trap and the main parameters

The magnetic field inside the trap is given by the following expressions:

$$B_x = \frac{4B_r}{\pi\sqrt{2}}\sum_n \frac{(-1)^n}{4n-3}\sin(k_n S)e^{-k_n d}\left(1-e^{-k_n D}\right)$$

$$B_y = \frac{4B_r}{\pi\sqrt{2}}\sum_n \frac{(-1)^n}{4n-3}\cos(k_n S)e^{-k_n d}\left(1-e^{-k_n D}\right) \qquad \text{(A)}$$

$$k_n = \frac{2\pi}{\lambda}(4n-3),\ \lambda = 2\ inches,\ D = 1\ inch.$$

(This magnetic field should be considered as an approximation of the real magnetic field created by small permanent magnets located at the bottom of the trap.) The explicit form of the function S is given below.

## 2. Equation of motion (adiabatic approach)

The equation of motion for a single neutron is:

$$\frac{d\vec{v}}{dt} = -\frac{\mu}{m}grad(B) + \vec{g}\ .$$

The magnetic moment of the neutron $\mu = 9.66236\times10^{-27}$ joules/tesla, the mass of the neutron $m = 1.67493\times10^{-27}$ kg, gravitational acceleration $g = 9.79 m/s^2$. Magnetic field $B = \sqrt{B_x^2 + B_y^2}$ .

$$\frac{\partial B}{\partial x} = \frac{B_x\frac{\partial B_x}{\partial x} + B_y\frac{\partial B_y}{\partial x}}{\sqrt{B}},\quad \frac{\partial B}{\partial y} = \frac{B_x\frac{\partial B_x}{\partial y} + B_y\frac{\partial B_y}{\partial y}}{\sqrt{B}},\quad \frac{\partial B}{\partial z} = \frac{B_x\frac{\partial B_x}{\partial z} + B_y\frac{\partial B_y}{\partial z}}{\sqrt{B}}.$$

$$\frac{\partial B_x}{\partial u} = +\frac{8B_r}{\lambda\sqrt{2}}\sum_n (-1)^n\left(1-e^{-k_n D}\right)e^{-k_n d}\left[\cos(k_n S)\frac{\partial S}{\partial u} - \sin(k_n S)\frac{\partial d}{\partial u}\right]$$

$$\frac{\partial B_x}{\partial u} = -\frac{8B_r}{\lambda\sqrt{2}}\sum_n (-1)^n\left(1-e^{-k_n D}\right)e^{-k_n d}\left[\sin(k_n S)\frac{\partial S}{\partial u} + \cos(k_n S)\frac{\partial d}{\partial u}\right]$$

The adiabatic approximation which is used in our simulation means that the magnetic moment of a neutron is oriented in the positive direction of the local magnetic field. The conditions of the validity of adiabatic approximation are analyzed in Sec. 5.

## Calculation of S and its derivatives

Introduce here the angle $\beta$ and the function S:

$$tg\beta = \frac{y_p}{a-u} = \frac{y_p}{\sqrt{(ra-z)^2 + x_p^{\ 2}} - r} = \frac{y_p}{raMu - r};\ S = y \text{ if } y_p = 0,\quad S = w/2 + a\times arctg\beta \text{ for } y_p > 0.$$

The explicit expressions for the derivatives of the function S, which are used in the numerical simulations, are:

$$\frac{\partial S}{\partial x} = \frac{-ay_p x_p}{\left(\sqrt{(ra-z)^2 + x_p{}^2} - r\right)^2 + y_p^2} \times \frac{sign(x)}{\sqrt{(ra-z)^2 + x_p{}^2}} = -\frac{y_p}{aMd^2} ax_p \times \frac{sign(x)}{raMu}.$$

$$\text{For } y_p > 0,\ \frac{\partial S}{\partial y} = \frac{a(raMu - r)\,sign(y)}{(raMu - r)^2 + y_p^2} = a\frac{aMu}{aMd^2} sign(y);\ \frac{\partial S}{\partial y} = 1 \text{ for } y_p = 0.$$

$$\frac{\partial S}{\partial z} = \frac{ay_p}{(raMu - r)^2 + y_p^2} \times \frac{(ra-z)}{raMu} sign(y) = a\frac{y_p}{aMd^2} \times \frac{(ra-z)}{raMu} \text{sign}(y).$$

## 3. Results of numerical experiments and optimization of parameters for a neutron trap

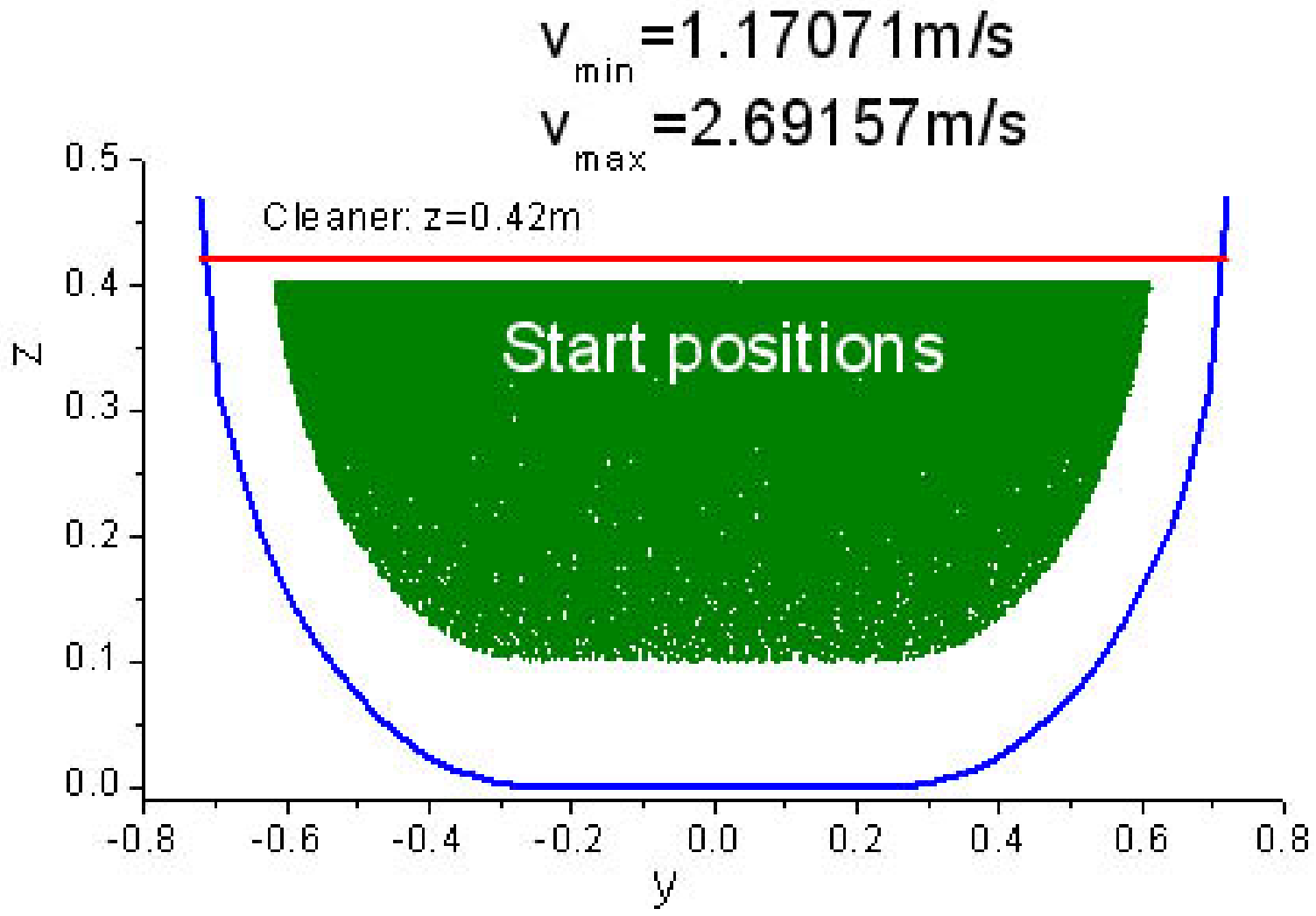

Fig. 1. The initial distribution of neutrons in the trap. $z_{\max} = 0.42m$, $h_{\max} = 0.47m$.

If we take into consideration only the first harmonic in the expressions for $B_x, B_y$, then the value of the magnetic field depends only on the distance to the surface, $d$. In what follows, we call this field a smooth field. Our first numerical experiments were done for smooth field:

$$B = 0.82\exp(-kd),\ \ k \approx 123.7m^{-1}.$$

Fig. 1 demonstrates the special distribution of neutrons at the initial time. The initial energy is defined by the maximal height which a neutron can reach $h_{\max}$, $E = mgh_{\max}$. For each neutron its initial coordinates in the central part of the trap were chosen randomly. After this, the initial kinetic energy was defined at a given point. The direction of the initial velocity was chosen randomly. The number of neutrons in our numerical experiments was varied from 100000 to 200000.

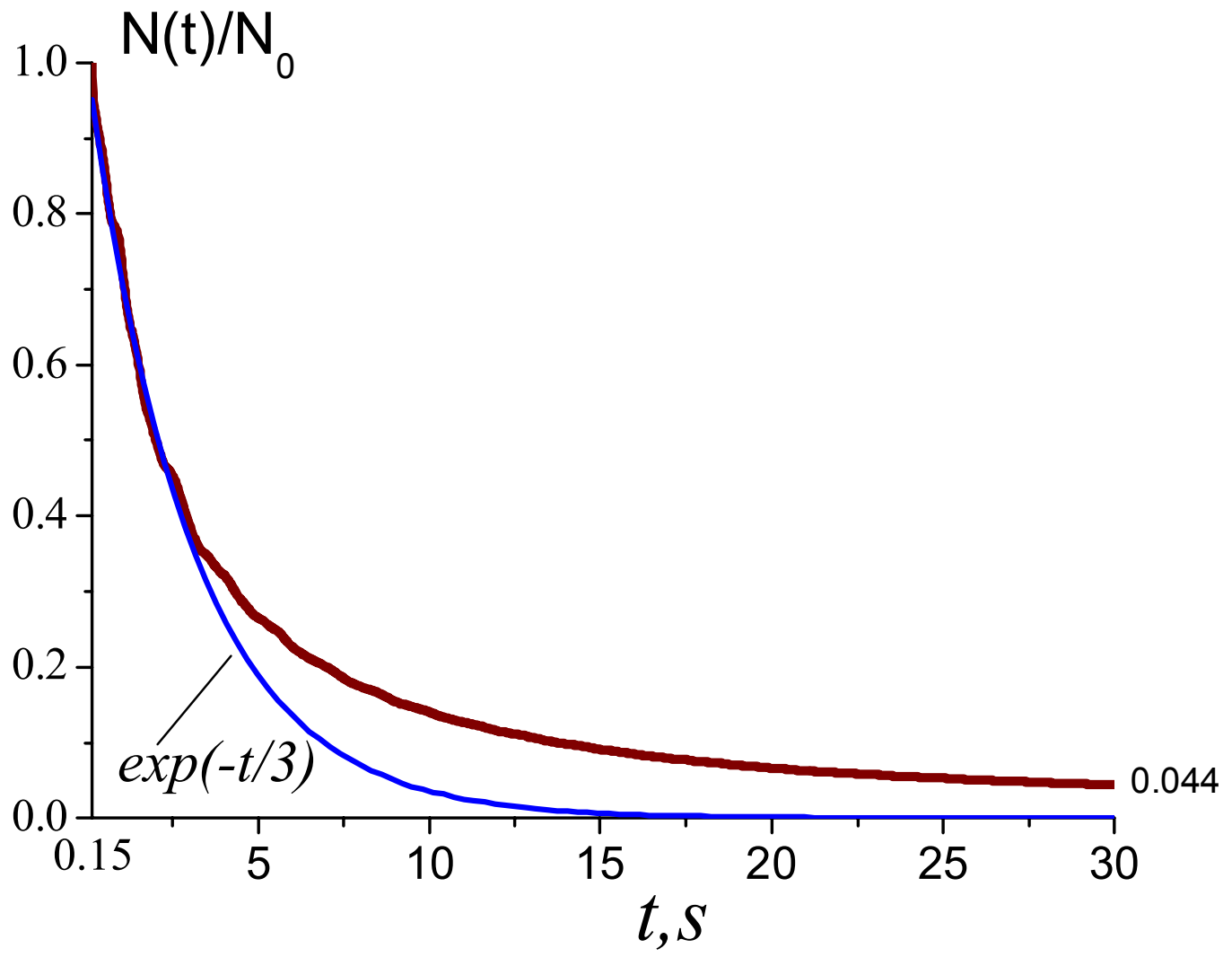

Fig. 2. The fraction of neutrons in the trap as a function of time (in the tail of the distribution), $N_0 = N(t = 0.15s)$. The case of the smooth magnetic field. $z_{max} = 0.42m$, $h_{max} = 0.47m$.

During the initial stage of the process (up to 0.15 sec) a rapid decrease in the number of neutrons takes place, related to the initial conditions: neutrons with large vertical components of velocity leave the trap. For the case, presented in Fig. 1, the number of neutrons decreases during this time from 200000 to 125000. After this time, the important process of cleaning the trap starts (Fig. 2).

In what follows, we usually will present not the whole dependence $N(t)$, but only the important part (the "tail") of this distribution.

As Fig. 2 shows, a relatively large number of neutrons still remain in the trap. These neutrons slowly leave the trap rather slow. This undesirable phenomenon can cause an error in the measurement of the neutron life time.

We consider two possible methods for improving of the parameters of the trap.

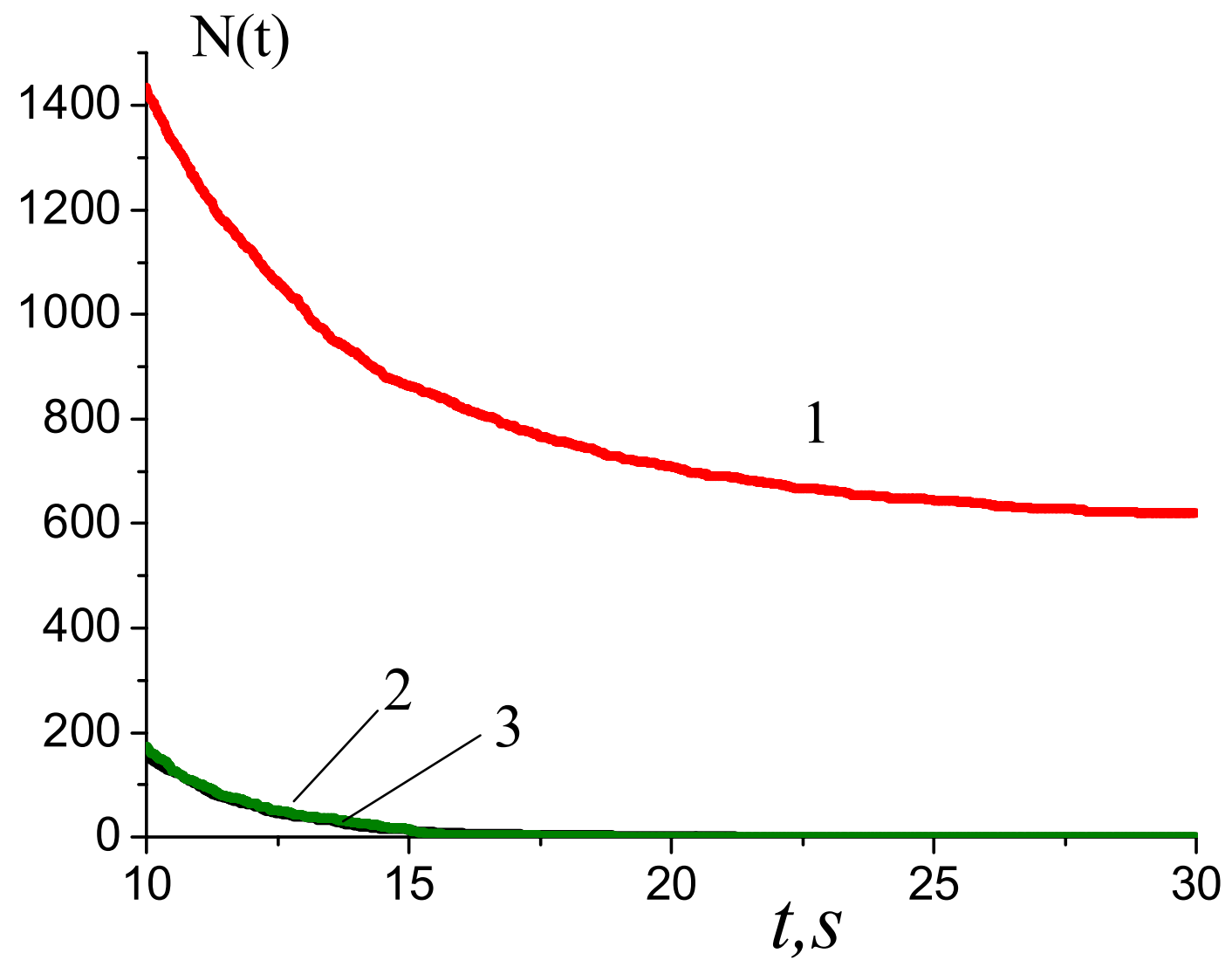

Fig. 3. The time-dependence of the number of neutrons in the trap under the mirror reflection ($z_{max} = 0.42m$, $h_{max} = 0.47m$). 1 – the diffuse reflection only from the curvilinear part of the trap; 2,3 – the diffuse reflection from the whole surface (2) and from only its flat part (3) (the dependences 2 and 3 are practically the same). The initial number of neutrons is 100000. If the surface reflects neutrons as a mirror, than $N(t = 30s) \approx 2800$. (See Fig. 5.)

**The first method: Formation of the diffuse reflection from the surface.** For the case of a smooth magnetic field, the surface of the trap acts as a "mirror" for the neutrons. Thus, the behavior of neutrons is analogous to behavior of tennis balls which rebound elastically are from the smooth surface. The escape of the neutrons from the trap can be accelerated if one assumes that the surface of the trap can act not as a mirror but as a diffuse one. The corresponding numerical experiments were carried out. We assumed that after approaching a minimal distance to the surface, the direction of the neutron's velocity changes randomly. The following cases were considered: (i) the whole surface of the trap is diffuse, (ii) only the curved part of the trap is diffuse, and (iii) only the flat part of the trap (bottom) is diffuse one. The results are shown in Fig. 3.

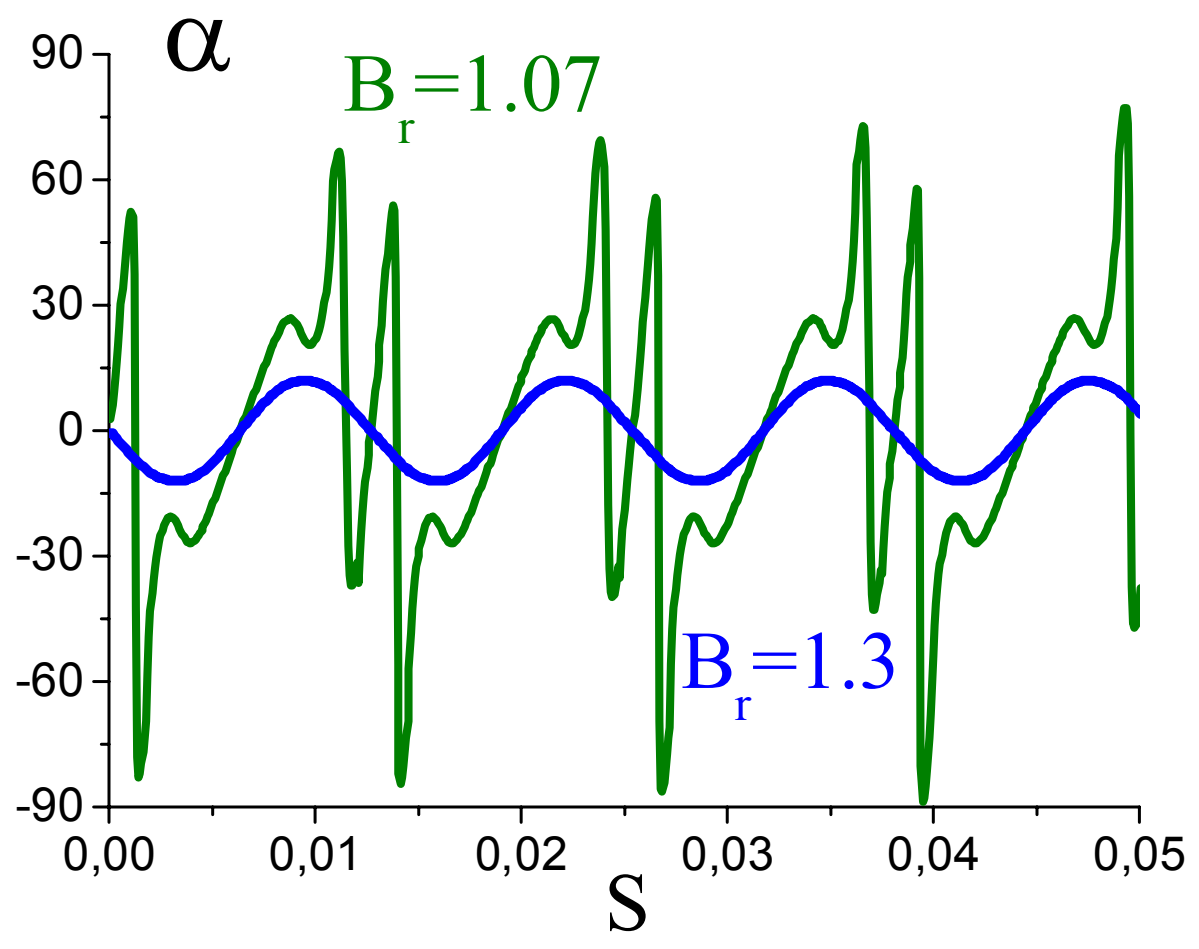

Fig. 4

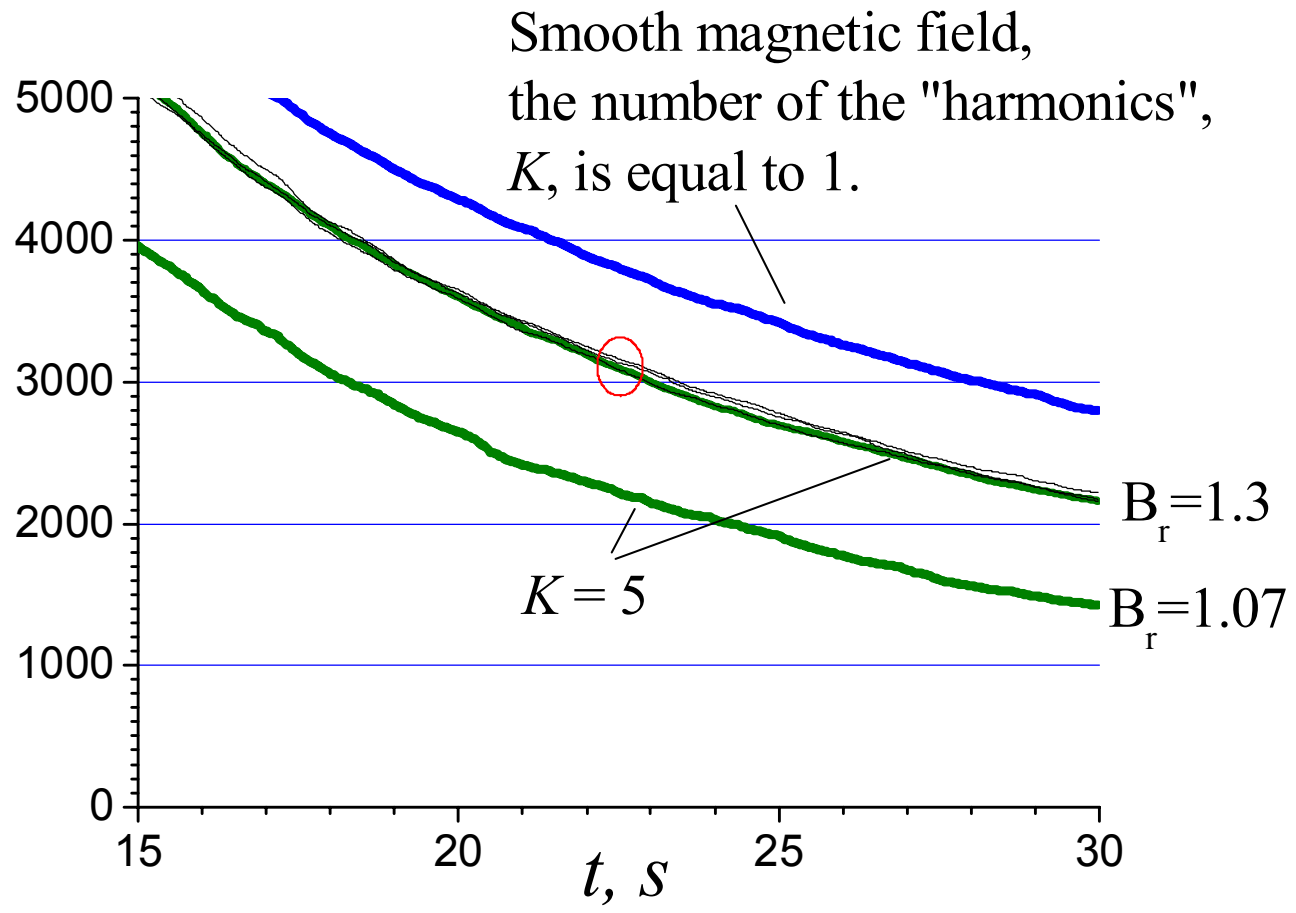

Fig. 5

Fig. 4. When decreasing the amplitude of the magnetic field, $B_r$, neutrons can penetrate closer to the surface, and the reflection becomes close to the diffuse one; $h_{\max} = 0.47m$.

Fig. 5. The "tail" of the dependency $N(t)$ for the different designs of the magnetic field. $z_{\max} = 0.42m$, $h_{\max} = 0.47m$, $N(t=0) = 100000$.

As demonstrated below, the role of the flat part of the trap varies – in some cases it improves the trap, in others it makes the situation worse. In the case shown in Fig. 3, the diffuse reflection is useful and provides a rapid escape for the untrapped neutrons. The diffuse reflection can be achieved by small-scale variations of the geometry of the surface of the bottom of the trap or by an inhomogeneous magnetic field. In this connection, we present our results for including in the expression for the magnetic field (A) the first five harmonics. Note, that the higher harmonics decrease rapidly with increasing distance from the surface. For revealing of high harmonics in neutron dynamics, the value of $B_r$ in (A) should be chosen in optimal way. If the value of $B_r$ is too small, then in some regions of the trap the neutrons could penetrate through the magnetic wall. If the value of $B_r$ is too large, then neutrons with total energy near the critical value, $E_c = mgz_{\max}$, can not approach the surface close enough to be affected by high harmonics.

For simplicity, we consider the case for the flat bottom of the trap. For a smooth magnetic field, its gradient is oriented in the direction of the normal to the surface in all points, providing a mirror-type reflection surface. If one takes into account the short-wave components of the magnetic field, then the angle $\alpha$ between the normal direction and the gradient of the magnetic field will change along the surface (Fig. 4). Let's demonstrate the variation of this angle for the motion of a neutron along the Y-axis, above the bottom of the trap. (According to (A) the modulus of the magnetic field changes just along the $Y$-axis.) The following fact must be taken into account. For a given total neutron energy, the minimal distance between this neutron and the surface, $d_{\min}$, is a function of $y$. That is why, the angle α in Fig. 4 is a complicated function of $y$: $\alpha = f\left(B_r\left(d_{\min}(y)\right)\right)$.

The data shown in Fig. 4 demonstrate that the reflection of the neutrons, at a given angle of approaching the flat surface, will take place at different angles, depending on the point contact with the magnetic wall. Up to some extent, this case is analogous to diffuse reflecting surface. Naturally, this effect is revealed more explicitly in weaker magnetic fields, when the neutron can approach the surface more closely. In this case, the speed of cleaning of the trap increases (see Fig. 5). Note, that for $B_r = 1.3$ the characteristics of the trap do not change if the number of harmonics in (A) exceeds 2. In this case, the neutron does not feel the short-wave components of the field, because it cannot move close enough to the wall.

It is possible to design a diffuse reflective bottom by covering it with granular magnetic particles. The parameters of the granules (or the regular edges on the surface) should be optimized using computer models.

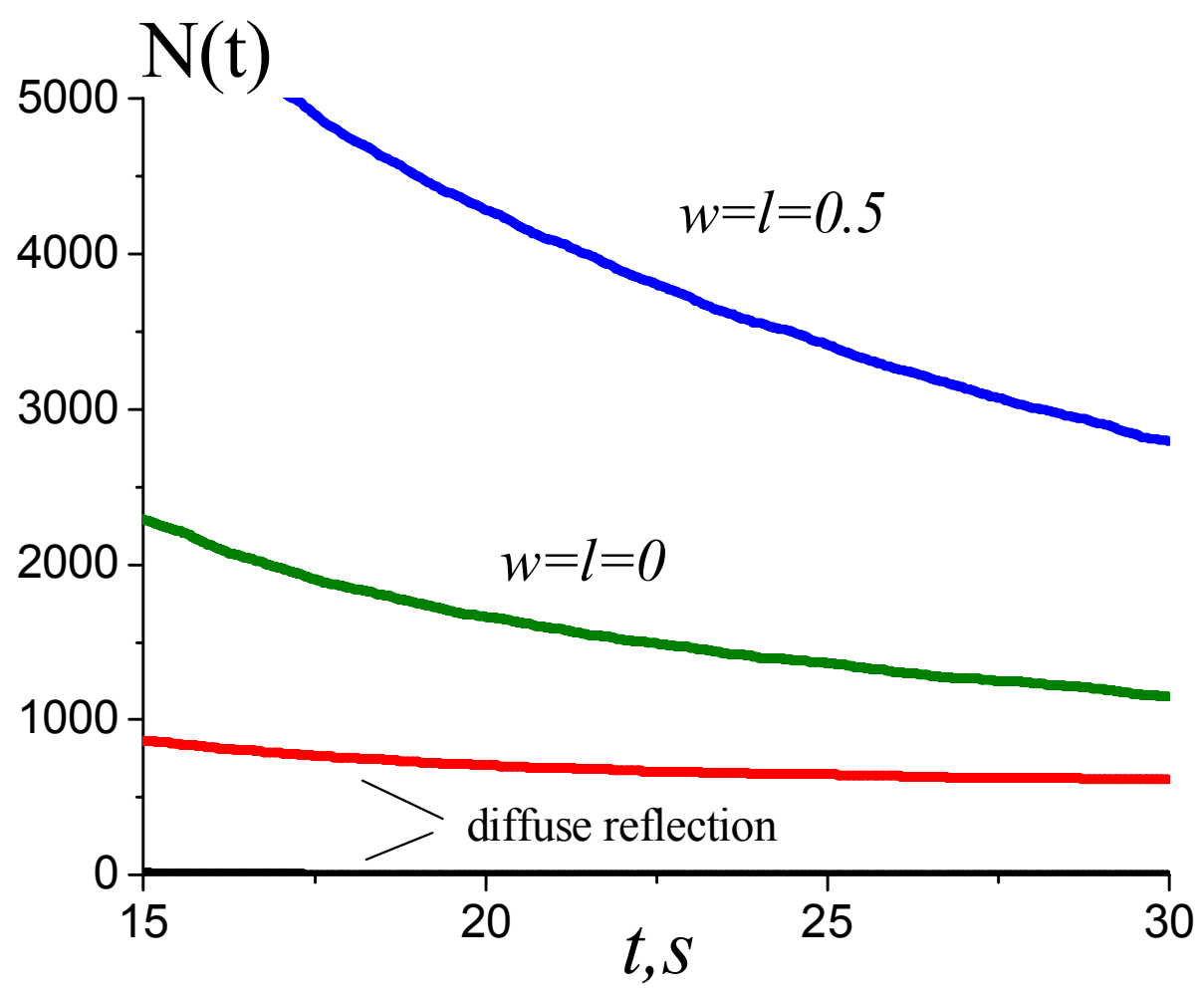

Fig. 6. Under the mirror reflection, the existence of the flat bottom decreases the escape of neutrons from the trap. The lower curves represent the results shown in Fig. 3. $z_{\max} = 0.42m$, $h_{\max} = 0.47m$, $N(t=0) = 100000$.

**The second case: The construction of asymmetric traps.** As numerical experiments demonstrate, the existence in the trap of the "mirror" flat bottom increases the time for cleaning of the trap (Fig. 6). This result requires a more thorough consideration in order to provide a qualitative and quantitative understanding. Here we present only the results of numerical simulations by letting the parameters of the trap, $w, l$, approach zero. (See Fig. 6, where a significant decrease of the cleaning time is demonstrated.) Thus, an important factor is the general shape of the trap. Namely, it should be asymmetric, and should not have a flat bottom. The first numerical simulations were done for the trap shown in Fig. 7. We will not describe here in detail its shape, as it appeared to be not the best variant of the asymmetric trap, but we present only the final results and compare them with the data presented in Fig. 5,6. At the end of 30 seconds, 900 neutrons were left in the trap. This result is better by about 30%, than for the symmetric trap without the flat bottom, but it still is not good enough for carrying out the experiments. Recall that we consider a system of neutrons with $h_{\max} = 0.47$. For smaller energies, $z_{\max} < h_{\max} < 0.47$, the escape time increases. The trap demonstrated in Fig. 7 is asymmetric and has a sharp back wall ($x < 0$) and a smoother front wall ($x > 0$).

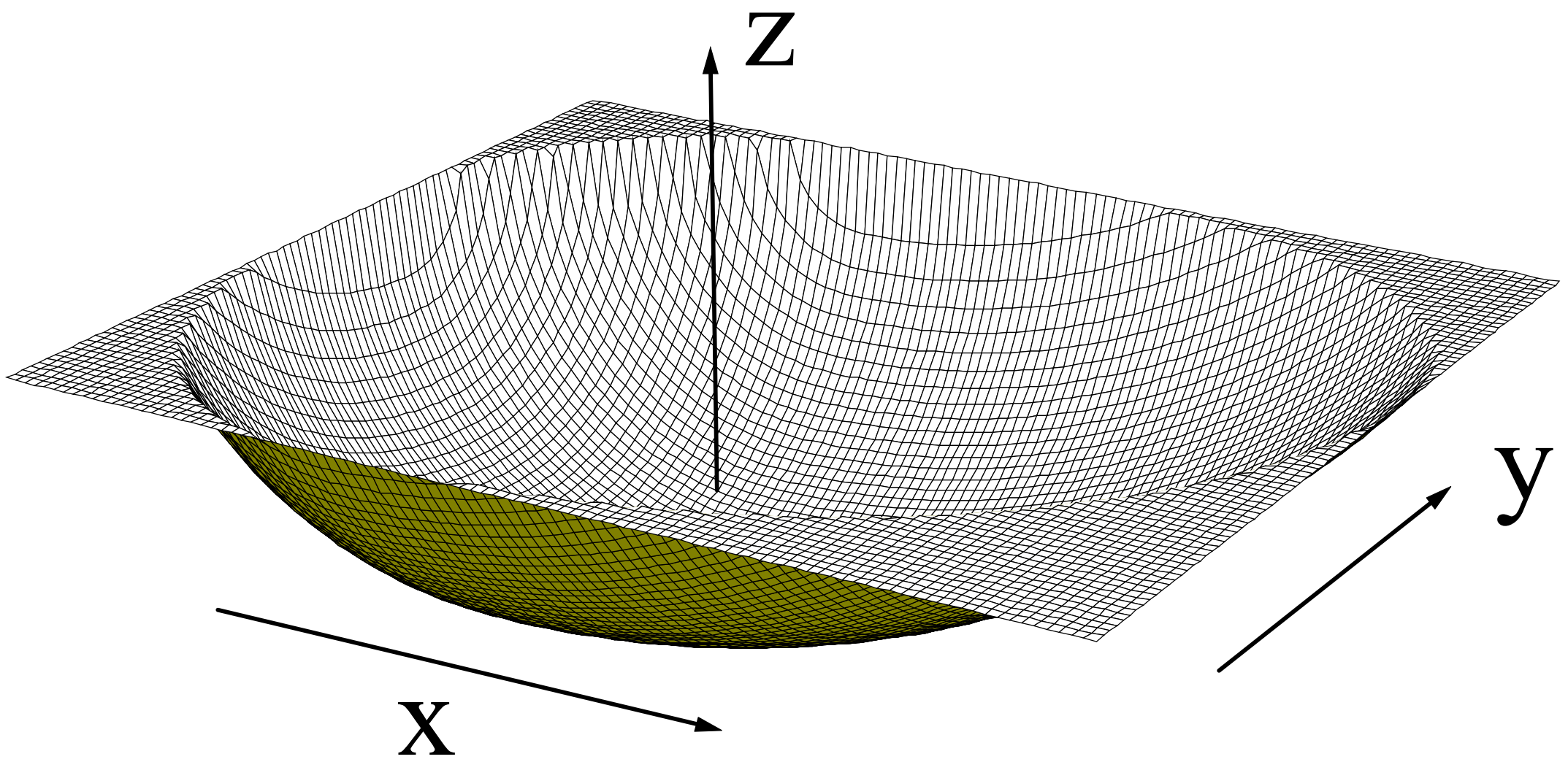

Fig. 7. One of the variants of the asymmetric trap.

Much better parameters demonstrated the trap which was designed from the trap shown in Fig. 7 in the following way. The closest to the reader part of the trap is reflected in the plane *yz* in the region $x < 0$. Thus, a "semi-symmetric" trap is realized when the symmetry relative to the plane *xz* is destroyed. The detailed description of the trap's form and its effectiveness are given below.

# 4. Asymmetrical trap and "smooth" magnetic field

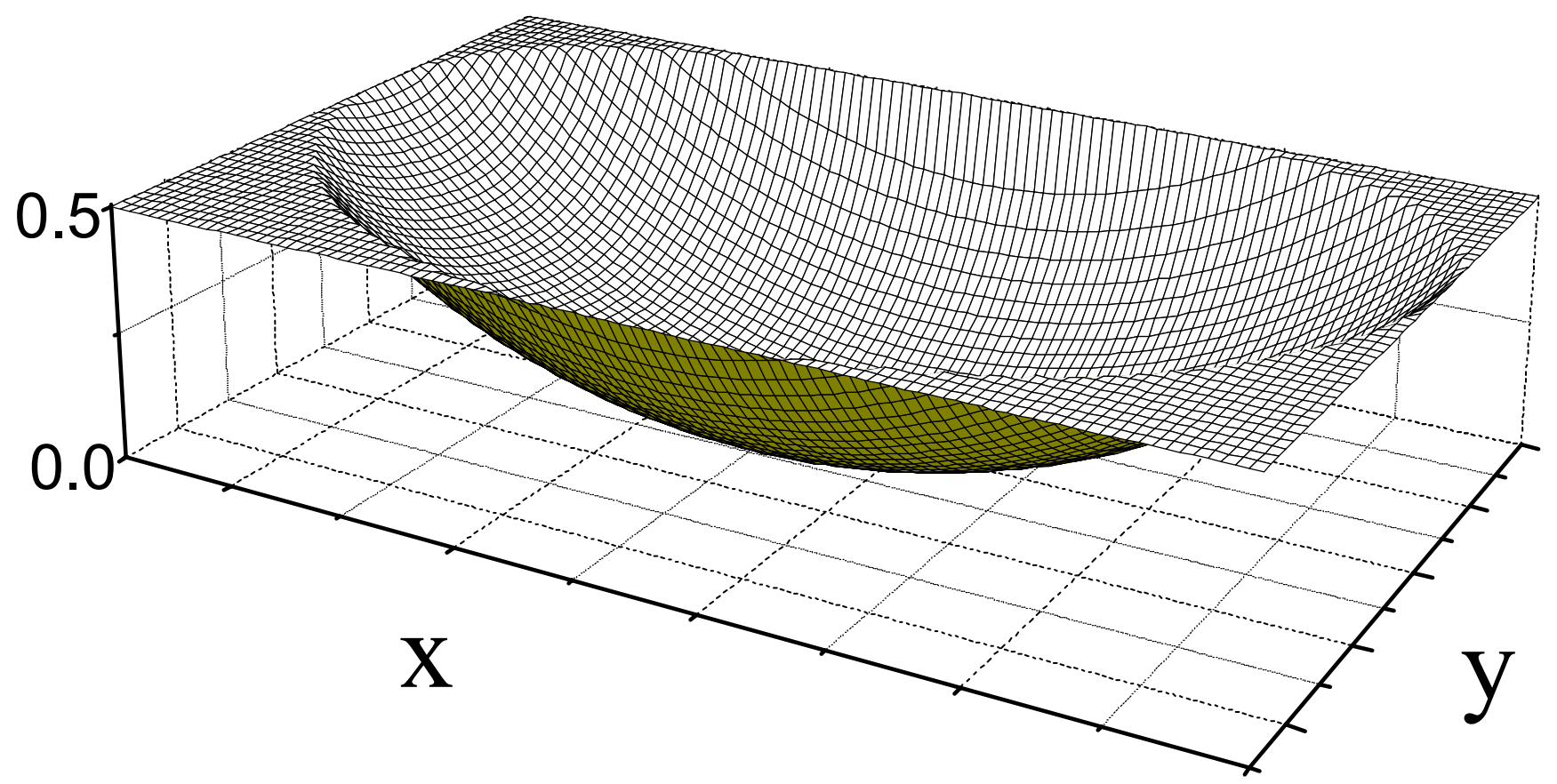

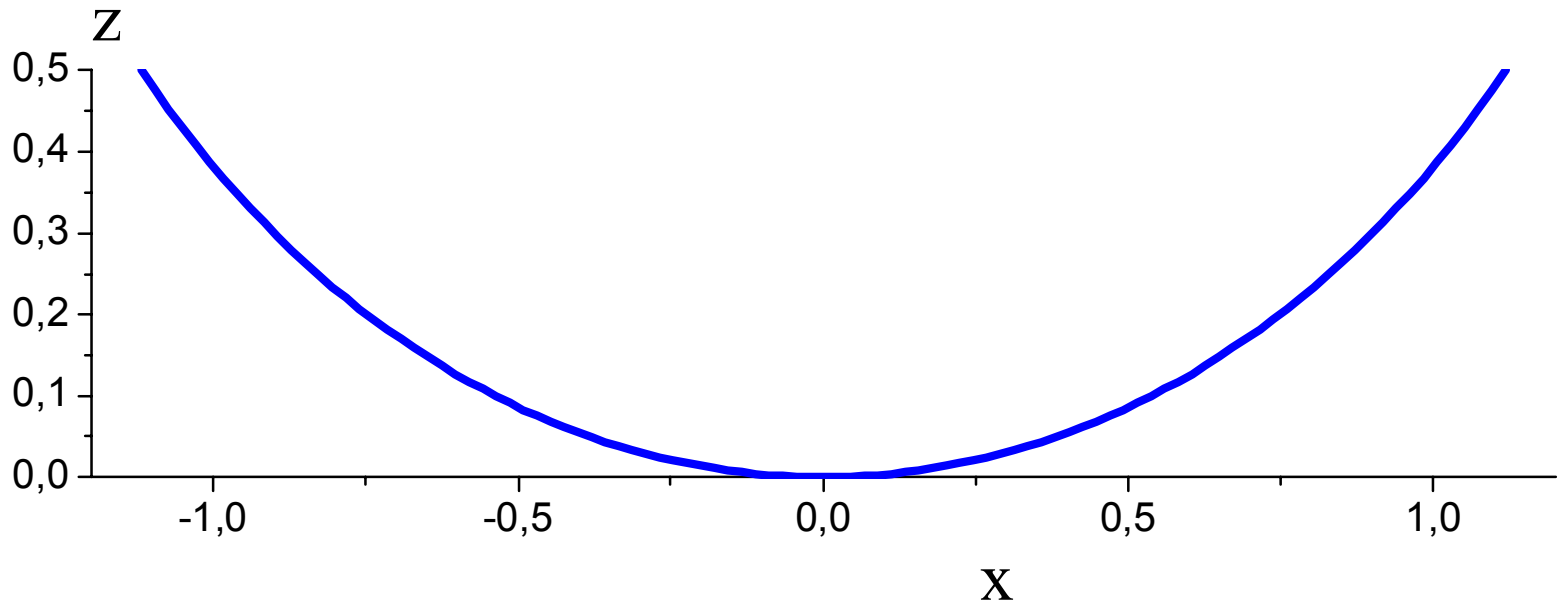

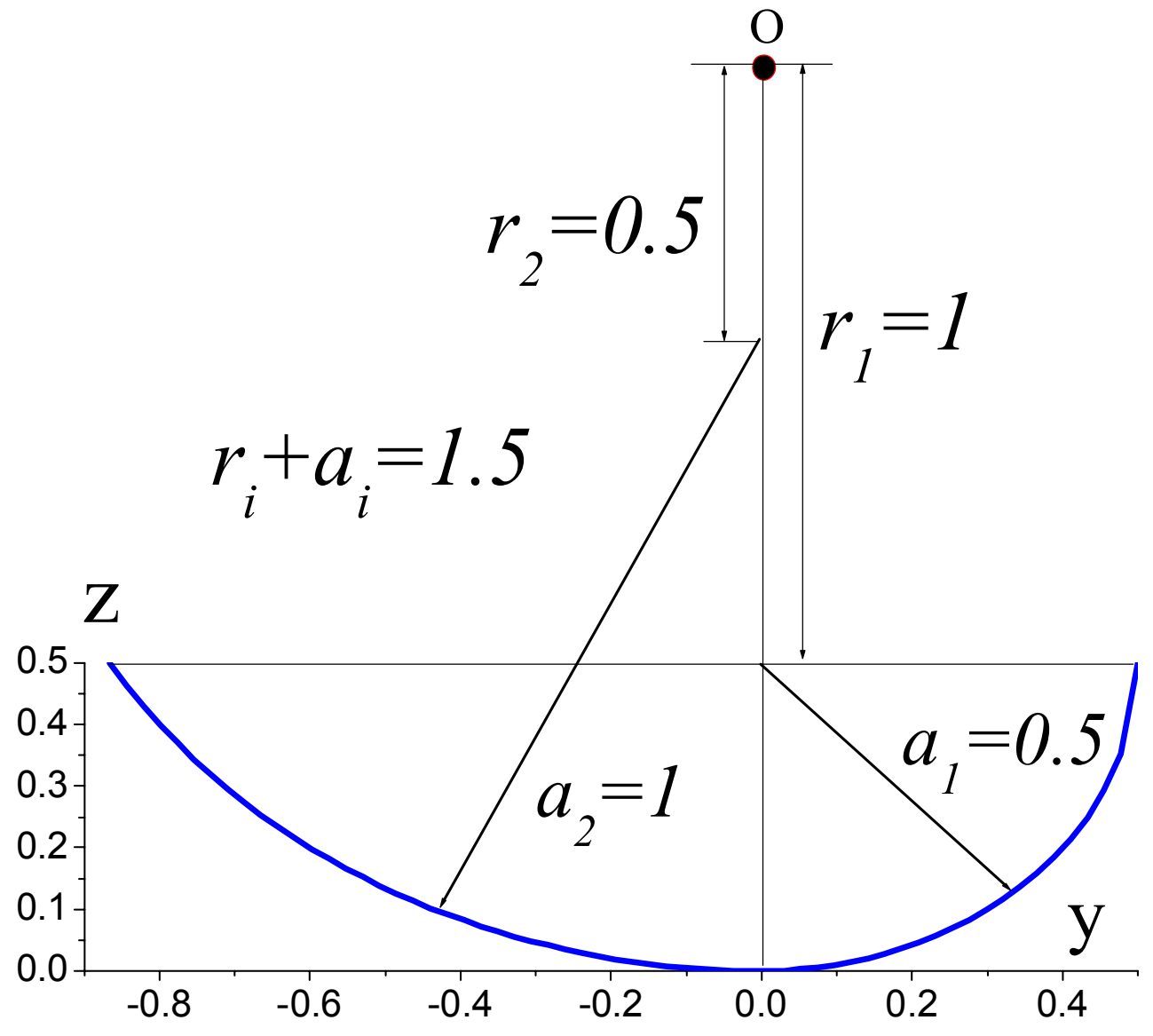

Fig.8. The shape of the trap which is asymmetric relative to the *y*=0 plane.

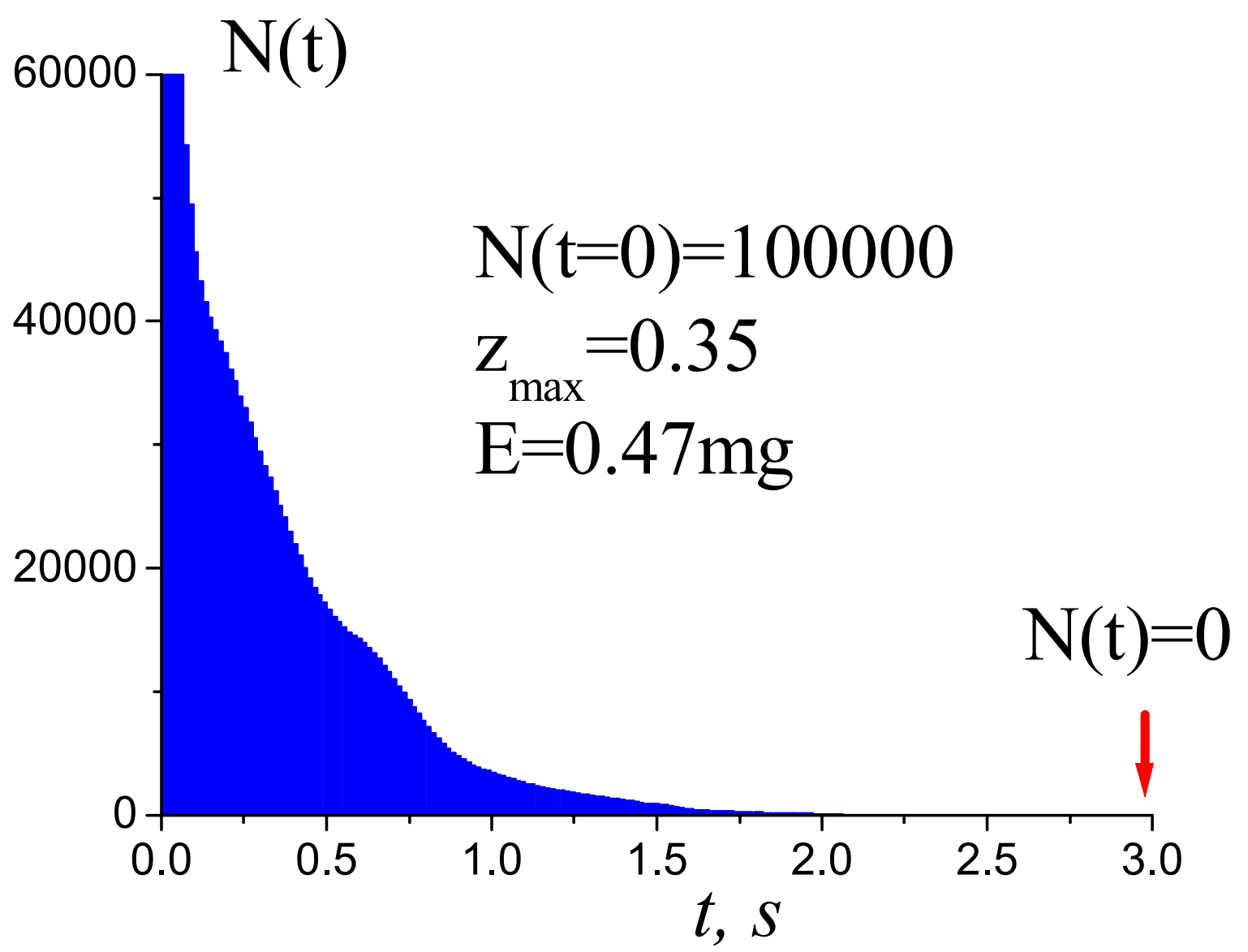

### *Results for the asymmetrical trap*

First, we assume that the neutron absorber is placed at $z_{max}$, which is significantly below the "maximal" height $h_{max} = 0.47m$. (See a figure on the left, in which $\Delta = h_{max} - z_{max} = 0.12$m.) Below, we present the results for neutrons with a smaller difference $\Delta = h_{max} - z_{max} \le 0.05$m. The asymmetric trap, even in these conditions, appeared to be most effective (see Fig. 9*a,b,c*) in comparison with the symmetric trap. (See Fig. 2 and Fig.5.)

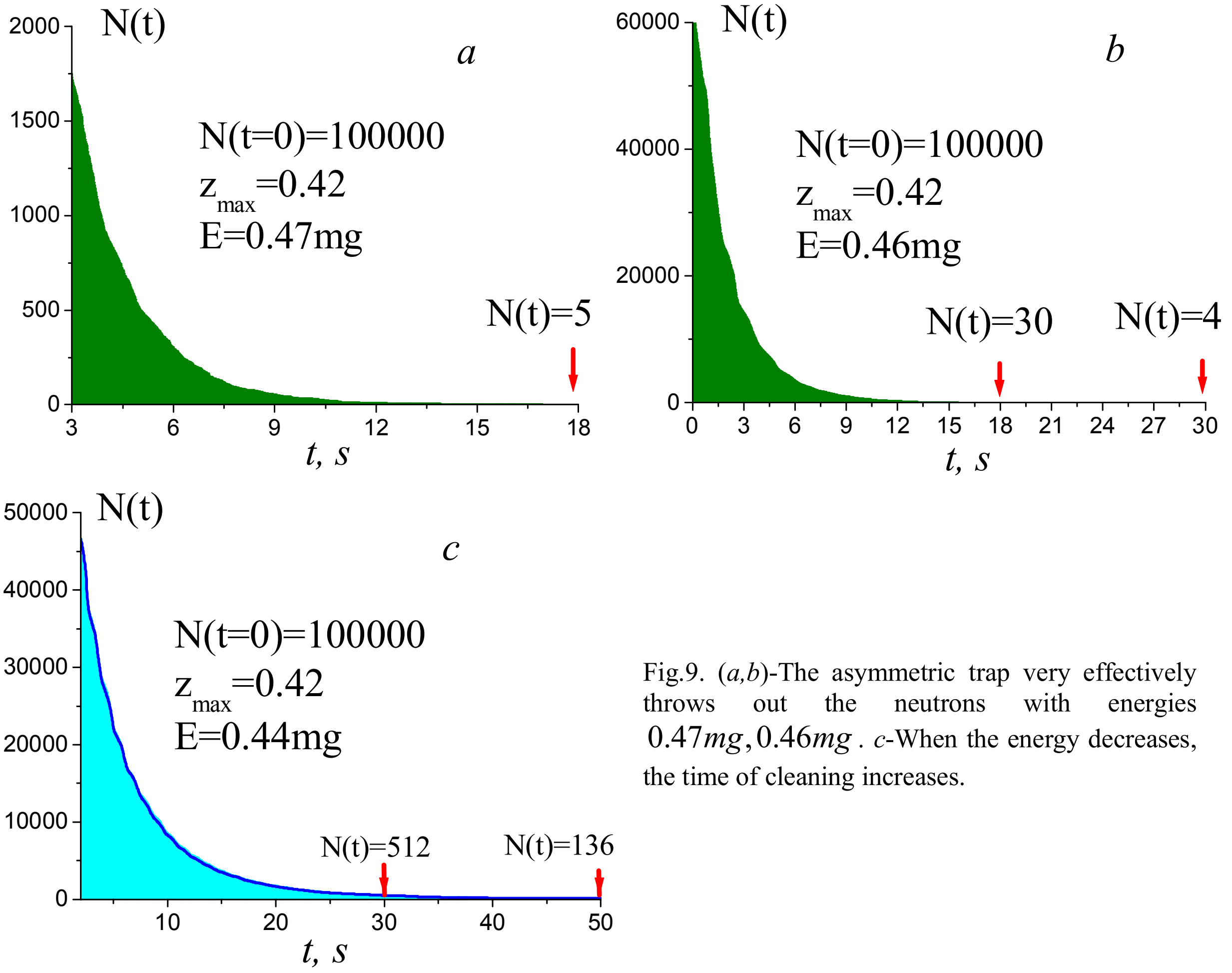

Fig.9. (*a,b*)-The asymmetric trap very effectively throws out the neutrons with energies $0.47mg, 0.46mg$. *c*-When the energy decreases, the time of cleaning increases.

To clean the trap for neutrons with energy $0.44mg$, we reduce the cleaner height to $z_{\max} = 0.4$. It appeared that the speed of cleaning is determined by the parameter $h_{\max} - z_{\max}$. Thus, the dynamics of $N(t)$, for parameters $z_{\max} = 0.4,\ h_{\max} = 0.44$ (see Fig. 10), appeared to be close to the dependence in Fig. 9b. After this change of the absorber height (from 0.42m to 0.40m), neutrons with the energy $E$=0.42$mg$ will escape. The difference $h_{\max} - z_{\max} = 2$ in this case is the same as in Fig. 9c. As a result, the curve $N(t)$ (Fig. 11) is close to the dependence in Fig. 9c.

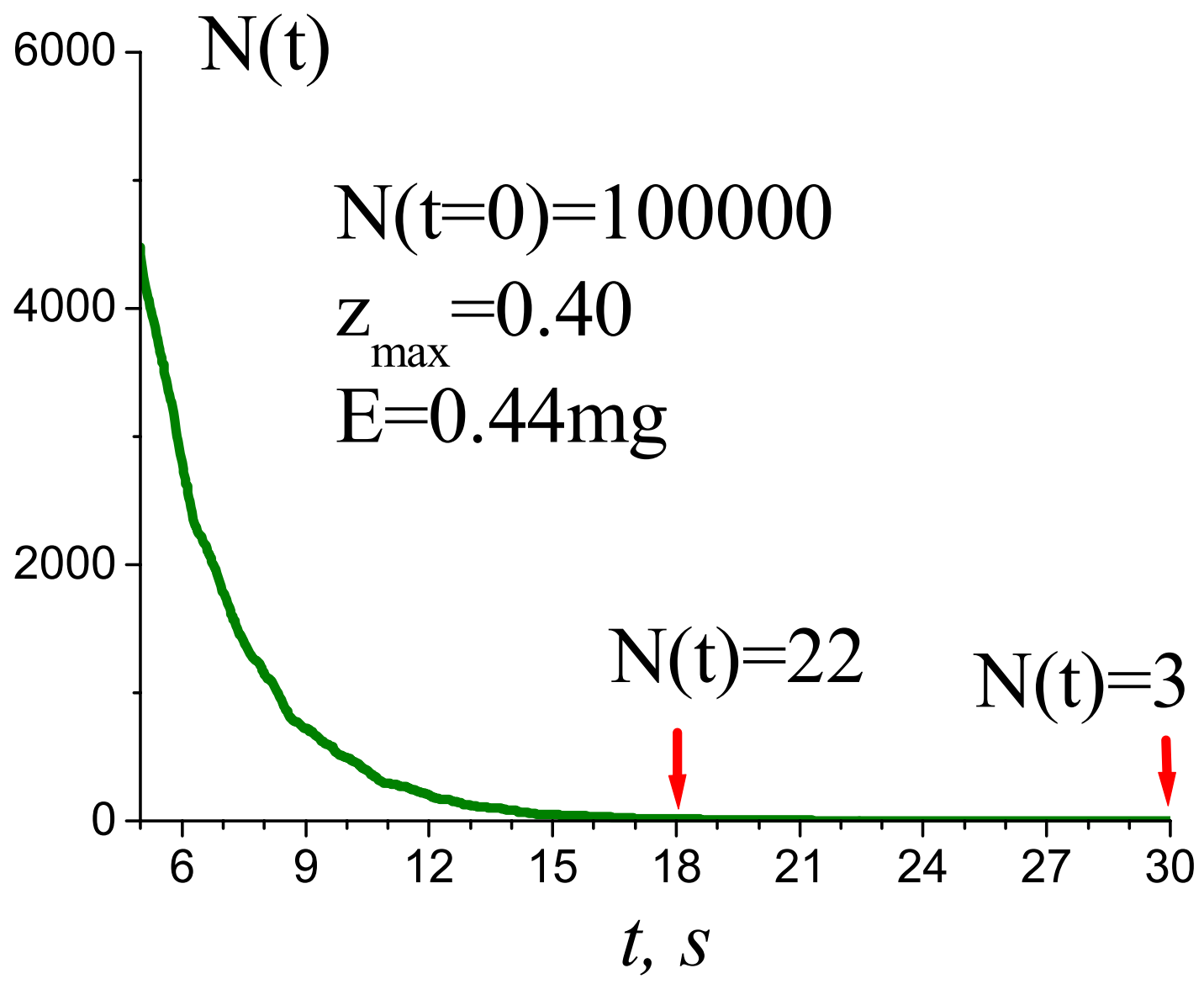

Fig. 10. The effective cleaning of the trap from neutrons with energy $E = 0.44mg$, reducing the height of the absorber (from 0.42m to 0.40m).

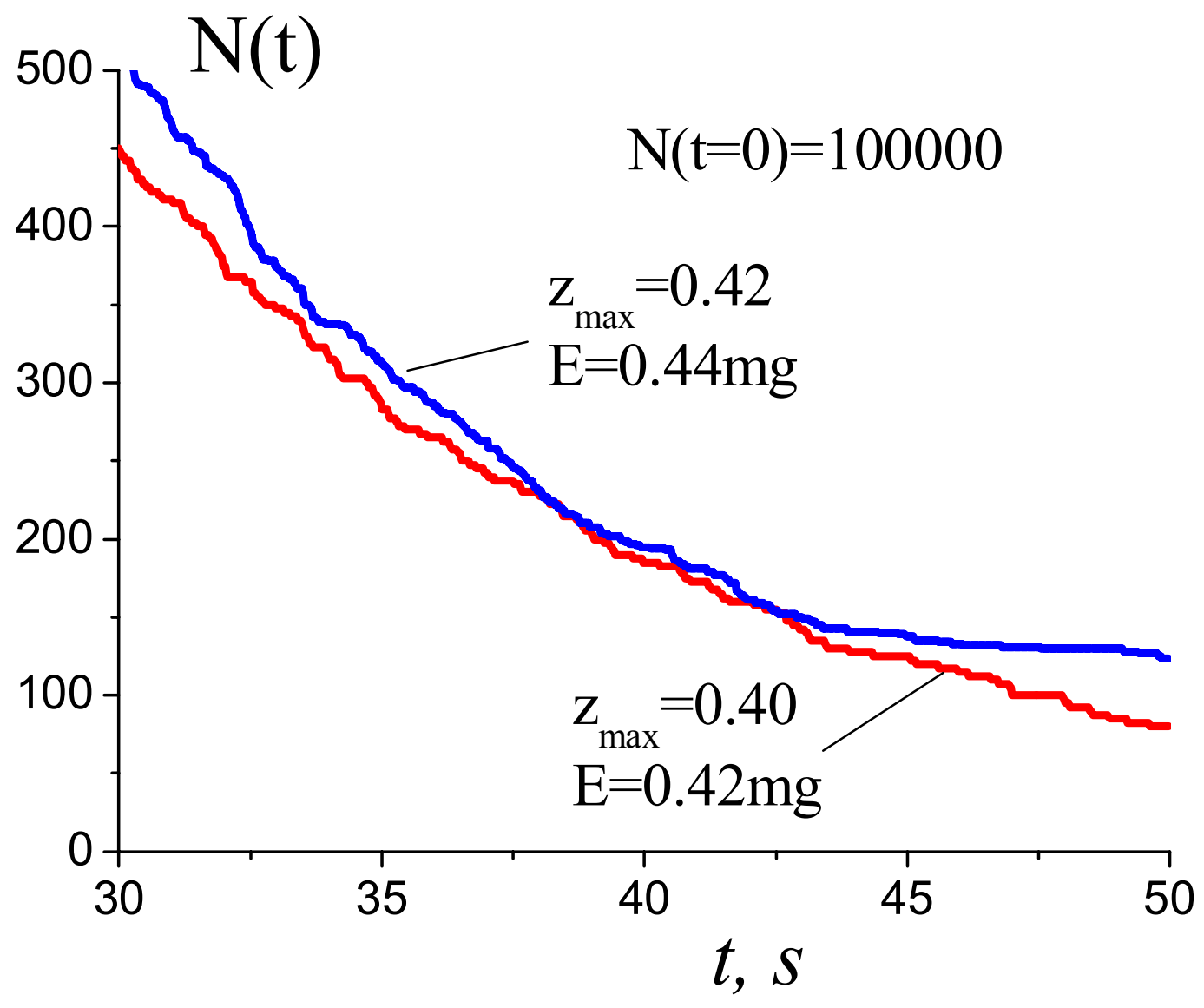

Fig. 11. Under the equal difference $h_{\max} - z_{\max}$, the curves $N(t)$ differ only by the statistical error.

**Supplement.** The effective cleaning of the trap from neutrons with energies close to the critical one can be achieved in, at least, two ways: (i) by optimizing the trap's shape, and (ii) by creating diffuse conditions for the diffuse reflection of neutrons. However, for the optimal shape of the trap the additional requirement of diffuse reflective walls is not evident. The numerical modeling demonstrated that the most effective diffuse reflection takes place from the flat bottom of the trap, not from the curvilinear walls (Fig. 3). For the chosen form of the trap (Fig. 8), which is probably close to the optimal one, the effect of the diffuse reflection does not change the cleaning time.

It could be that the most effective solution of the problem corresponds to the intermediate region, when both the form of the trap and the diffuse reflection should be incorporated. The solution of this problem requires a thorough theoretical and numerical analysis.

***Taking into account the electron guide field:*** $B_\phi = B_e \dfrac{r+a}{\rho}$. ($B_{tot} = \sqrt{B_x^2 + B_y^2 + B_\phi^2}$)

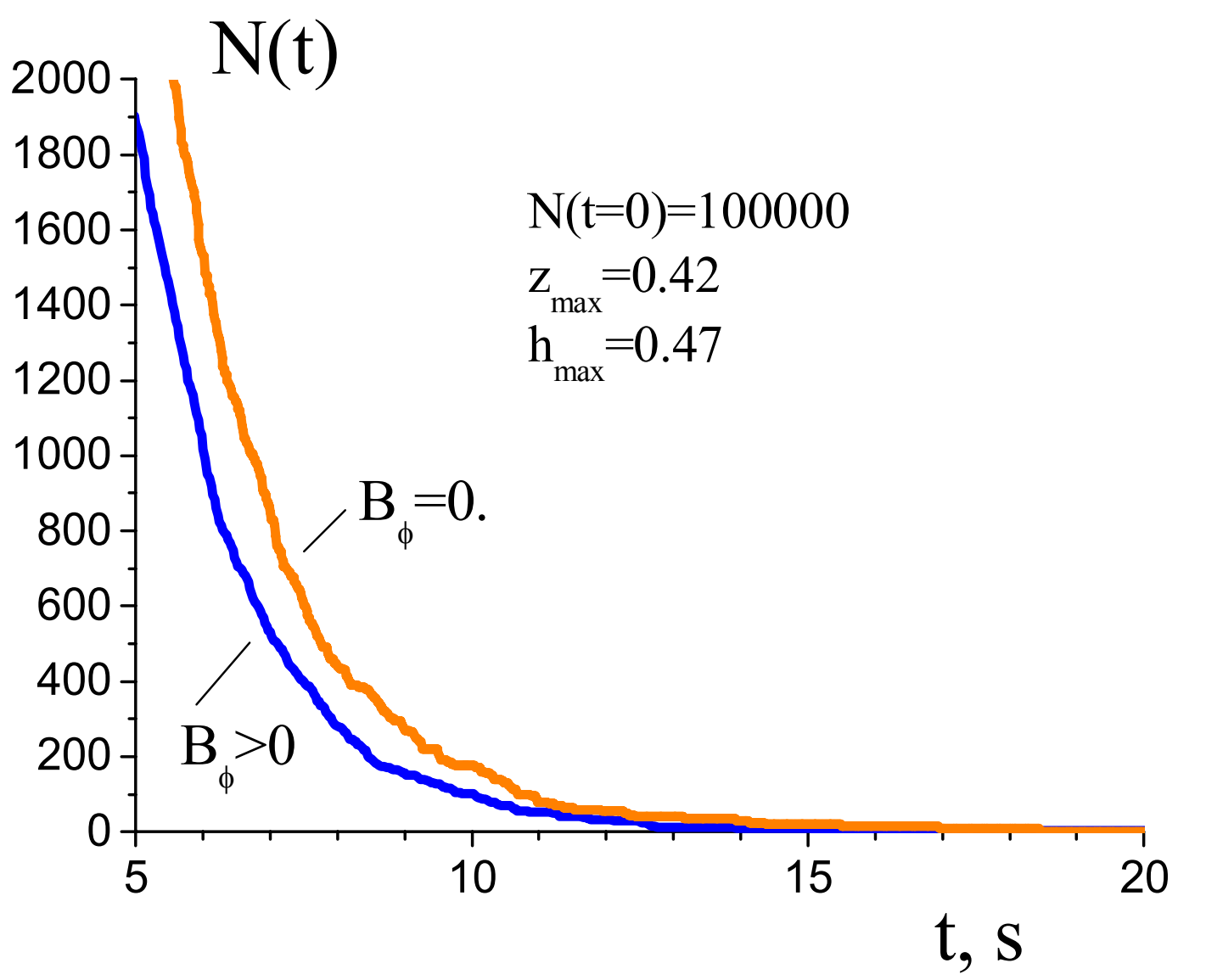

Fig. 12. The comparison of the escaping speeds of neutrons for $B_\phi = 0$ and $B_\phi > 0$. The existence of the weak magnetic field in the trap does not make the characteristics of the trap worse.

$\rho$ - is the distance from the point $O$ in Fig. 8 to the projection of an arbitrary point in the space on the plane $xz$. At the value $B_e = 0.05T$ the electron guide field increases from $0.05T$ in the lower part of the trap to $\sim 0.07T$, near the absorber. The initial total energy of a neutron, for a given value of $h_{\max}$, should include the energy due to the electron guide field:

$$E = h_{\max} mg + \mu B_e \frac{r+a}{r+a-h_{\max}}.$$

This value corresponds to the total energy of a neutron located at $h_{\max}$, above the center of the trap ($x = 0, y = 0$).

## 5. Conditions for adiabatic dynamics

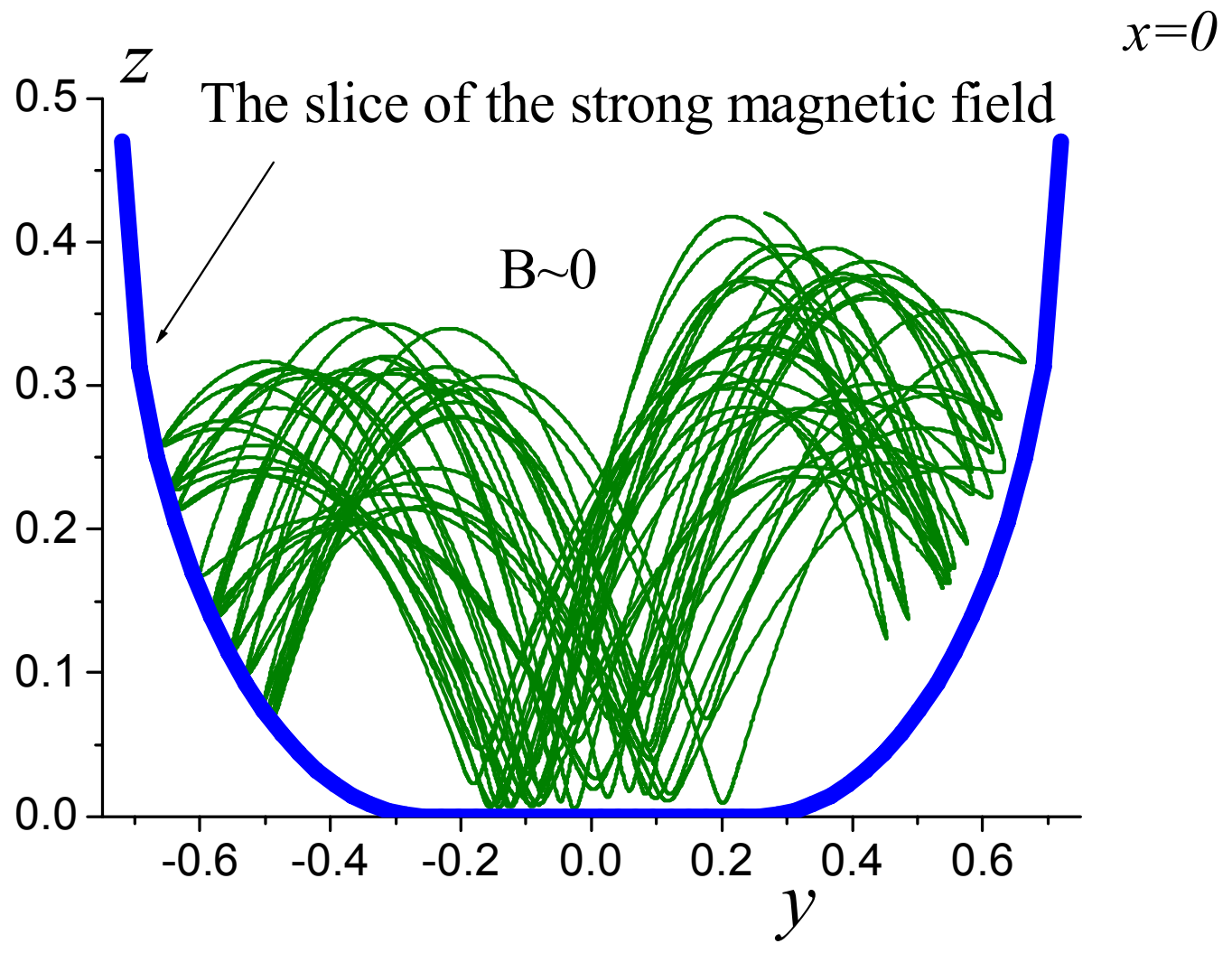

Fig.13. Untrapped trajectory of a neutron obtained by using the adiabatic invariant. To provide the adiabatic dynamics, a low external magnetic field $B_0$ was applied, $B_0 = 3 \times 10^{-3} T$.

The conditions required for adiabatic dynamics for neutron spin are reduced to the condition that the value of the Larmor frequency of a neutron around the local magnetic field is larger than the characteristic frequency of variation of the magnetic field in the space due to motion of the neutron. Consider the case in which the condition of adiabaticity, in regions far from the surface, is provided by a weak magnetic field, $B_0$, directed along the $Z$-axis. In this case, a spin experiences a large change in its orientation: from the vertical to the horizontal one in the vicinity of the flat bottom, Fig.14. (We consider the case of a symmetric trap; a part of the characteristic trajectory of a trapped neutron is shown in Fig. 13; five harmonics of the magnetic field are taken

into account.)

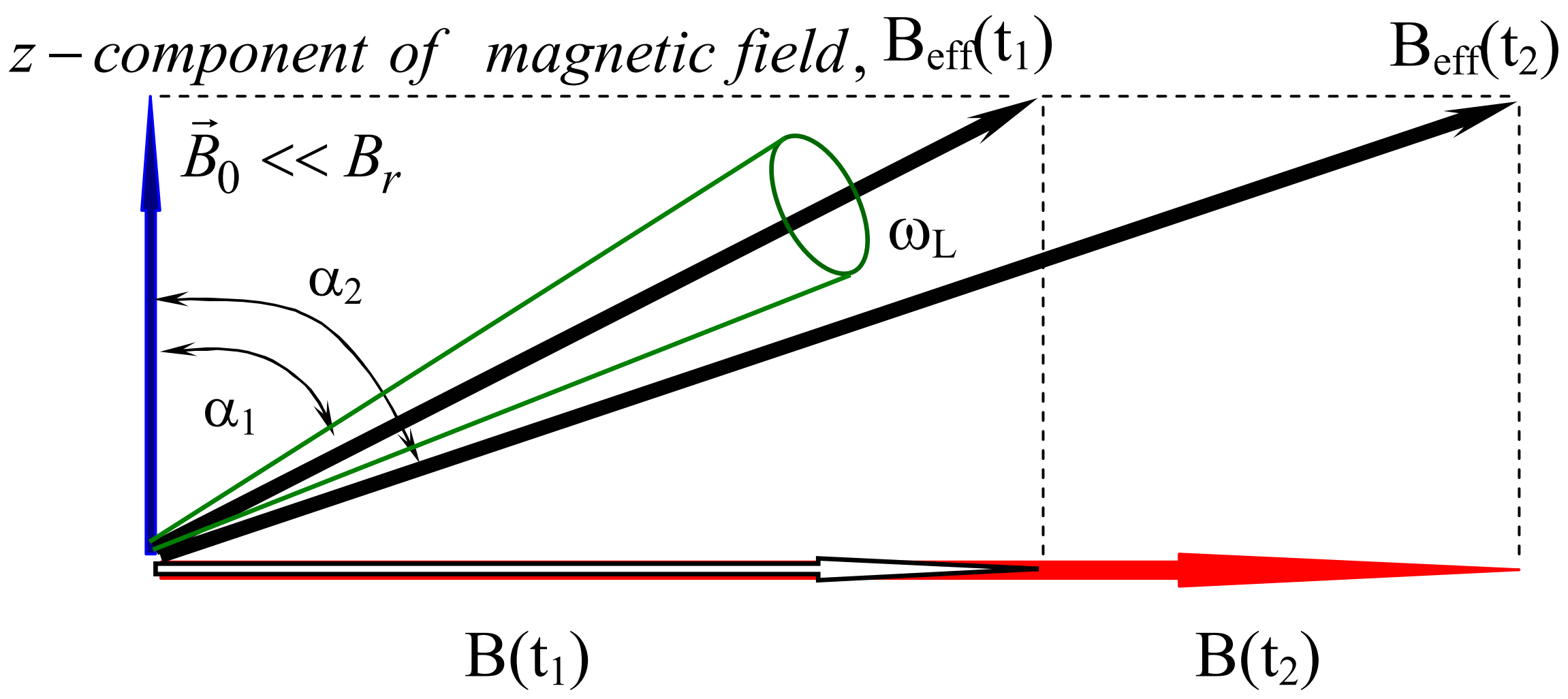

Fig.14. Illustration of the rotation of the neutron spin as it approaches the reflecting wall. $\omega=(\alpha_2-\alpha_1)/\Delta t$-the frequency of the variation of the magnetic field in the vicinity of the surface (collision with the wall).

For the adiabatic invariant approach to be valid, the inequality $|\omega(t)| << \omega_L = \gamma_n B_{eff}(t)$ has to be satisfied ($\gamma = 1.832 \cdot 10^8 s^{-1} T^{-1}$). The Larmor frequency in our case is not less than the Larmor frequency in the magnetic field $B_0$: $\omega_L(t) \ge \omega_L^{(0)} = \gamma_n B_0 > 5 \cdot 10^5 s^{-1}$. The frequency of variation of the magnetic field as a function on time is shown in Fig. 15 for the trajectory shown in Fig. 13.

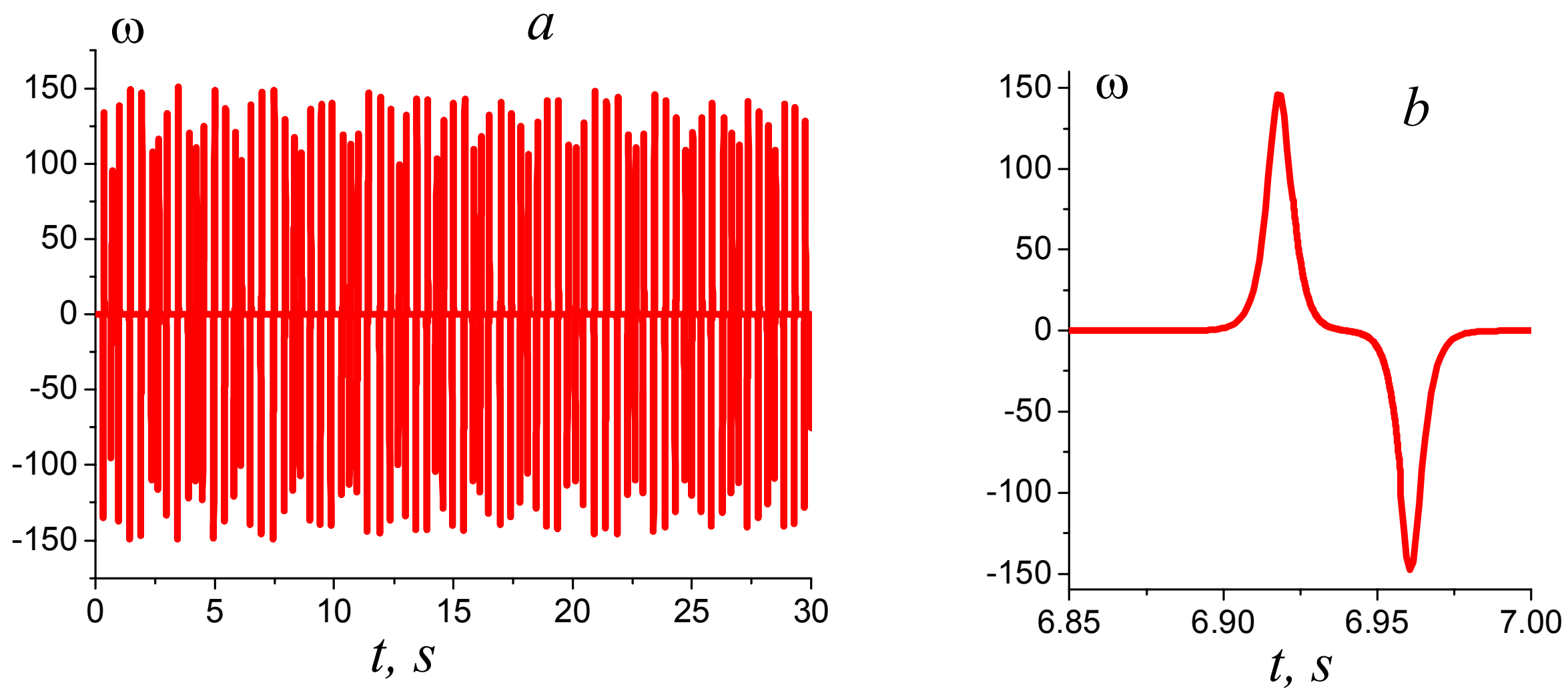

Fig. 15: *a* - The dependence $\omega(t)$ for the trajectory presented in Fig. 13. b - The dynamics of $\omega(t)$ during a single collision with the wall: $\omega_{\max} \approx 150 s^{-1} \ll \omega_L^{(0)}$.

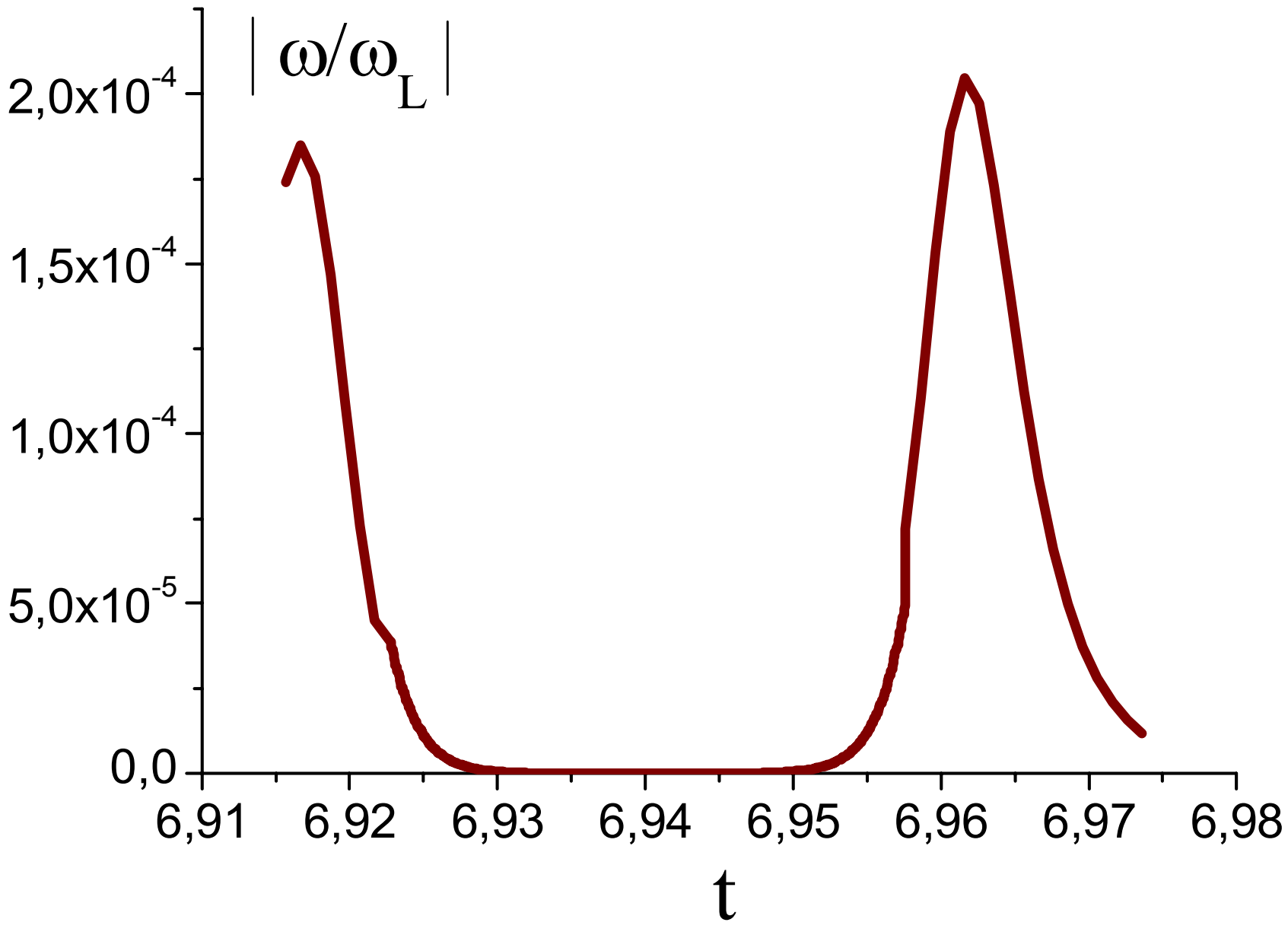

Fig. 16. The dynamics of the ratio $|\omega / \omega_L|$ for the time interval shown in Fig. 14*b*.

The curves shown in Figs. 15 and 16 justify the validity of the adiabatic approximation. The value of the magnetic field $B_\phi$ is larger than one used in our simulations with the constant magnetic field: $B_0 = 0.003T$.

In conclusion, the adiabatic invariant approach is valid even for a sufficiently low enough external magnetic field $B_0$.

## Acknowledgement

We are thankful to David Bowman (P-23, NEUTRON SCIENCE & TECHNOLOGY) and Peter Walstrom (LANSCE-ABS: LANSCE ACCELERATOR, BEAM & SPALLATION PHYSICS) for suggesting us these simulations and for many useful discussions.